\documentclass[twocolumn,superscriptaddress,aps,preprintnumbers,amsmath,amssymb,footinbib,prl]{revtex4}
\usepackage{graphicx}
\usepackage{dcolumn}
\usepackage{bm}
\usepackage{hyperref}
\usepackage{times}
\usepackage[dvips]{color}

\newcommand{\be}{\begin{equation}}
\newcommand{\ee}{\end{equation}}
\newcommand{\bea}{\begin{eqnarray}}
\newcommand{\eea}{\end{eqnarray}}

\begin{document}
\title{Topological Dark Matter}

\author{Hitoshi Murayama}
\affiliation{Institute for the Physics and Mathematics of the
  Universe, University of Tokyo, Kashiwa 277-8568, Japan}
\affiliation{%
  Department of Physics, University of California, Berkeley, CA 94720}
\affiliation{%
  Theoretical Physics Group, Lawrence Berkeley National Laboratory,
  Berkeley, CA 94720 }
\author{Jing Shu}
\affiliation{Institute for the Physics and Mathematics of the
  Universe, University of Tokyo, Kashiwa 277-8568, Japan}

\begin{abstract}
  Kibble mechanism drastically underestimates the production of
  topological defects, as confirmed recently in atomic and condensed
  matter systems.  If non-thermally produced, they can be cosmological
  dark matter of mass 1--10 PeV. If thermalized, skyrmion of mass
  1--10 TeV is also a viable dark matter candidate, whose decay may
  explain $e^\pm$ spectra in cosmic rays recently measured by PAMELA,
  FERMI, and HESS. Models that produce magnetic monopoles below the
  inflation scale, such as Pati--Salam unification, are excluded.
\end{abstract}

\date{\today}
\maketitle
\preprint{UCB-PTH-09/14}
\preprint{IPMU09-0058}

Topological defects are of common interest to condensed matter
physics, atomic physics, astrophysics and cosmology, as well as
algebraic topology \cite{Anderson}.  When the symmetry group $G$
spontaneously breaks down to its subgroup $H$, there are continuously
connected ground states parametrized by the coset space $G/H$.  The
homotopy groups of the coset space then tell us what kinds of
topological effects are possible.  In most cases, non-trivial
$\pi_d(G/H)$ implies the existence of $(2-d)$-dimensional topological
defect.  If the coset space has disconnected pieces ($\pi_0 (G/H) \neq
0$), we expect {\it domain walls}\/.  For multiply-connected space
($\pi_1 (G/H)\neq 0$), there are {\it strings}\/ ({\it vortices}\/).
If the boundary of space can map non-trivially to the coset space
($\pi_2(G/H)\neq 0$), we expect point-like defects such as {\it
  magnetic monopoles}\/.  An exception to the rule is when the whole
space is mapped non-trivially to the coset space ($\pi_3(G/H) \neq
0$), where {\it skyrmions}\/ are stabilized by non-renormalizable
terms in the low-energy effective theory \cite{Skyrme:1961vr}.  In
this case, it is not the boundary condition that is topologically
non-trivial, but the configuration in the bulk.

To estimate the initial abundance of defects produced by a phase
transition in early universe, Kibble pointed out that the correlation
length diverges at the critical temperature while the causality does
not permit exchange of information beyond the horizon scale
\cite{Kibble:1976sj}.  He therefore came up with a {\it lower bound}\/
on the amount of defects, namely approximately one per horizon, called
Kibble mechanism.  Most of the literature uses this lower bound as the
estimate of the abundance of topological defects from phase
transitions in early universe.  For point-like topological defects,
one finds $n_{TD}/s \sim (T_c/M_{Pl})^3$.  Therefore, only phase
transitions close to the grand-unification scale produce abundance of
topological defects worthy of consideration.

A decade later, Zurek \cite{Zurek:1985qw} proposed a more refined
estimate of the abundance by carefully considering the time scale
available.  His estimate has been confirmed experimentally in a large
number of systems recently, now called Kibble--Zurek mechanism.  The
studies include liquid crystals \cite{Chuang,Bowick}, superfluid
$^4$He \cite{Hendry} and $^3$He \cite{Ruutu,Baeurle}, an optical Kerr
medium \cite{Ducci}, Josephson junctions \cite{Carmi,Monaco},
superconducting films \cite{Maniv}, and spinor Bose--Einstein
condensate \cite{Sadler}.

We point out that the Kibble--Zurek mechanism provides a substantially
larger abundance of topological defects from phase transitions in early
universe than the original estimate by Kibble.  Therefore even phase
transitions just above the TeV energy scale may produce interesting
(or dangerous) amount of topological defects.

In particular, we discuss the possibility that point-like topological
defects may be the cosmological dark matter, which is arguably one of
the most pressing mysteries in cosmology, astrophysics, and particle
physics \cite{Murayama:2007ek} \footnote{Our estimate on the initial
  density produced by a second order phase transition applies to
  non-point-like topological defects, such as cosmic strings and
  domain walls.}. The dominant paradigm to explain the dark matter is
the thermal relic of yet-undiscovered particle.  Within this paradigm,
we consider dark matter candidates below approximately 100~TeV in mass
because of the unitarity bound \cite{Griest:1989wd}. Our main result
in this Letter is that the natural range for topological dark matter,
if non-thermally produced by a second-order phase transition, is
$O(1\sim10)$~PeV, which obviously violate the unitarity limit. Note
that a symmetry breaking at this energy scale in the hidden sector is
of great interest in many attempts to understand the origin of
hierarchy between the Planck and electroweak scales such as dynamical
supersymmetry breaking, and extra dimensions. In addition, we also
point out that skyrmions at the order 10 TeV, once thermalized, are
also interesting dark matter candidates that are often ignored in the
literature \cite{baryons}. The existence of skyrmion solution is quite
generic in models where Higgs serves as a pseudo Nambu Goldstone
boson, which opens the new possibility to connect the origin of
electroweak symmetry breaking and dark matter.

If the dark matter particles are produced thermally at temperatures
higher than their mass, their initial abundance is the same as any
other relativistic particle species.  Then the final abundance is
determined by their annihilation cross section,
\begin{equation}
  \label{eq:OmegaX}
  \Omega_X h^2 \approx \frac{1.1 \times 10^9 (\ell+1) x_f^{\ell+1} {\rm
      GeV}^{-1}} {g_*^{1/2} M_{Pl} \langle \sigma v_{\it rel}\rangle_f}
  \approx \frac{3\times 10^{-27} {\rm cm}^3/{\rm sec}}{\langle \sigma
    v_{rel}\rangle_f}
\end{equation}
where $x_f = m/T_f$ with $T_f$ the freeze-out temperature, and we used
$g_* \approx 100$ and $\ell=0$ ($S$-wave).  Assuming that only one
partial wave $J$ would contribute, the annihilation cross section is
limited from above by \cite{Griest:1989wd}
\begin{equation}
  \label{eq:unitarity}
  \sigma_J v_{\it rel} < \frac{4\pi(2J+1)}{m^2 v_{\it rel}}
  \approx \frac{3\times 10^{-22}(2J+1) {\rm cm}^3/{\rm sec}}{(m/{\rm
      TeV})^2}\ .
\end{equation}
Combining Eqs.~(\ref{eq:OmegaX},\ref{eq:unitarity}), we find $m <
110$~TeV assuming $S$-wave annihilation and $J=0$.

On the other hand, the Kibble--Zurek mechanism predicts a very
different abundance of point-like topological defects.  Throughout
this paper, we assume second-order phase transition.  The correlation
length $\xi$ and relaxation time $\tau$ diverge near the critical
temperature which can be parametrized using the critical exponents
$\nu$ and $\mu$ respectively
\begin{equation}
  \label{eq:critical}
  \xi = \xi_0 | \epsilon |^{- \nu} \ ,  \qquad
  \tau = \tau_0 | \epsilon |^{-\mu} \ ,
\end{equation}
where $\epsilon \equiv (T_c - T)/T_c$ characterizes the proximity to
the critical temperature $T_c$.

The system is {\it quenched}\/ when it passes through the critical
temperature with a finite speed.  It is characterized by the quenching
rate $\tau_Q \equiv (t-t_c)/ \epsilon$ to the linear order around time
$t_c$ when $T=T_c$. During the quenching, there exists a particular
time $t_{*}$ when the time remaining before the transition equals the
equilibrium relaxation time $|t_*-t_c| = \tau(t_*)$. Beyond this point
the system can no longer adjust fast enough to follow the changing
temperature of the bath, and at time $t_*$ the fluctuation becomes
frozen until a time $|t_*-t_c|$ after the critical temperature is
reached.  It is easy to see that $|\epsilon(t_*)| =
(\tau_Q/\tau_0)^{-1/(1+\mu)}$.  Therefore, the fluctuation does not
get smoothed out beyond the correlation length \footnote{The Hubble
  expansion can be ignored when the system is close to the critical
  point during the phase transition as we can see that $1/H$ is much
  longer than the frozen relaxation time $\tau(t_*) \sim \tau_0
  (\tau_Q / \tau_0) ^{\frac{\mu}{1+\mu}}$.}
\begin{equation}
  \xi(t_*) \sim \xi_0 (\tau_Q / \tau_0) ^{\frac{\nu}{1+\mu}} \ .
\end{equation}
In radiation dominated universe, $T \propto t^{-1/2}$ and one finds
$\tau_Q = 2 t_c= H(T_c)^{-1}$ with the expansion rate $H=\dot{a}/a$.

Assuming the free energy of the Landau--Ginzburg form $V(\phi) = (T -
T_c) m \phi^2 + \frac{1}{2} \lambda \phi^4$ near $T_c$, one can
approximate $m \sim \lambda T_c$ and $\xi$, $\tau$ scale as $\xi_0 /
\sqrt{\epsilon}$, $\tau_0 / \sqrt{\epsilon}$ classically.  So the
critical exponents are $\mu = \nu = \frac{1}{2}$.  Setting the initial
correlation length $\xi_0 \approx \tau_0 \sim 1/(\sqrt{\lambda} T_c)$,
we have
\begin{equation}
  \xi \approx \left(\frac{T_c}{H} \right)^{1/3} \frac{1}{\lambda^{1/3} T_c} 
  = H^{-1} \left( \frac{H^2}{\lambda T_c^2} \right)^{1/3} \ . 
\end{equation}
In radiation dominated universe 
\begin{equation}
  H =  \frac{T^2}{C M_{pl}} \ , \qquad C = \sqrt{\frac{45}{4\pi^3 g_*}}
\end{equation}
and hence the correlation is shorter than the horizon size by a factor
$\sim (T_c/M_{Pl})^{2/3}$, leading to a far larger number of defects
than the original Kibble's estimate.

For point-like defects (PD), we expect approximately one per $\xi^3$.
Assuming ($g_{*} \approx 10^2 - 10^3$, $\lambda \approx 0.3$--1), we
find
\begin{equation}
  \left. \frac{n_{PD}}{s}\right|_{T=T_c} \approx 0.1 \frac{T_c}{M_{pl}} \ .
\end{equation}

If we consider the quantum corrections to the system, the critical
exponents $\mu$ and $\nu$ could be different from 1/2. Generally
speaking, it is related with the anomalous dimension of the leading
relevant operators in the of the Lagrangian of the scalars that
triggers the symmetry breaking.  Causality $\xi \leq c \tau$ dictates
$\nu \leq \mu$ and we will assume $\nu=\mu$ below as the Hubble
friction term for scalar field $\phi$ could be ignored in the vicinity
of critical temperature indicated in footnote \cite{endnote40}.  For
typical quantum systems based on $O(N)$-symmetric $\phi^4$ theory in
three dimensions \footnote{The symmetry breaking pattern $O(N)/O(N-1)$
  here does not have a non-trivial second homotopy group which leads to
  monopoles. Nevertheless, we take those well tested examples as
  illustrations.}, the critical exponents are $\nu = 0.625$ in binary
liquid system ($N=1$), $\nu = 0.672$ in superfluid $^4\textrm{He}$
experiment ($N=2$), and $\nu = 0.70$ in EuO, EuS system ($N=3$). As we
can see, $\nu$ is quite close to 2/3 and does not vary very much for
different $N$. By plugging in the same numbers as the classical case,
we obtain
\begin{equation}
\label{eq:monopoleratio}
  \left. \frac{n_{PD}}{s}\right|_{T=T_c} \approx 0.006 \left(
  \frac{30 T_c}{M_{pl}} \right)^{\frac{3\nu}{1+\nu}} \ .
\end{equation}

We see that the magnetic monopoles are produced orders of magnitude
more than the original Kibble's estimate and hence even models with
phase transitions down to TeV scale are subject to serious
constraints. For instance, Pati--Salam model \cite{Pati:1974yy}
assumes the symmetry breaking $SU(4)_C \times SU(2)_L \times SU(2)_R
\rightarrow SU(3)_C \times SU(2)_L \times U(1)_Y$, and hence predicts
magnetic monopoles.  Once produced, the monopoles are stable and their
number can only be reduced by annihilation of $M-\bar{M}$ pairs. The
magnetic (hidden) monopoles will stay in \textsl{kinetic} equilibrium
with the (hidden) thermal plasma of electrically charged
particles. The long-range forces between $M$ and light charged
particles will dissipate the energy of $M$ drifting towards a nearby
$\bar{M}$, allowing capture and annihilation to occur.  Preskill
\cite{Preskill:1979zi} found the annihilation was negligible assuming
the initial abundance given by the Kibble mechanism.  With the
Kibble--Zurek mechanism, however, the annihilation must be considered
for magnetic monopoles, and we must use Eq.~(5) in
\cite{Preskill:1979zi},
\begin{equation}
  \label{eq:monopoannihi}
  \frac{n_M}{T^3}  = \frac{1}{Bh^2} \left(\frac{4 \pi}{h^2} \right)^2
  \frac{m_{PD}}{C M_{pl}} \approx 7.9 \times 10^{-22} \times \left(
    \frac{T_c}{1\textrm{TeV}} \right) \ ,  
\end{equation}
where $B = (3/ 4 \pi^2) \zeta(3) \sum_i (h q_i/ 4 \pi)^2$ which sums
over all spin states of relativistic charged particles and $h=2 \pi
/q$ is the magnetic coupling. 

\begin{figure}
  \begin{center}
    \includegraphics[width=0.47\textwidth]{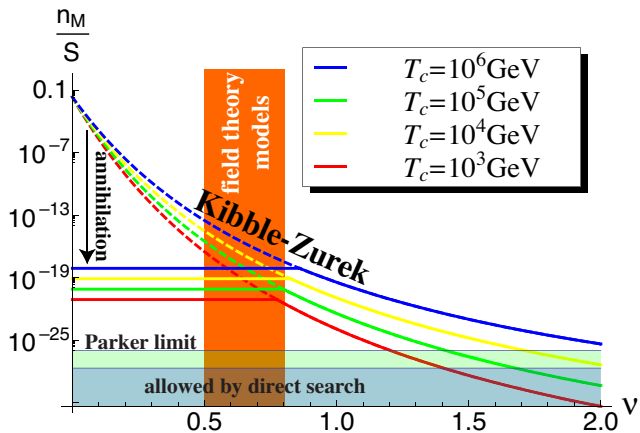}
    \caption{$n_M/s$ versus the critical exponent $\nu$ if the
      magnetic monopoles are 
      produced by a 2nd order phase transition.  The dashed line are
      the $n_M/s$ without the annihilation. The green region is
      allowed by the Parker's bound $n_M/s \lesssim 10^{-26}$ while
      the blue region is allowed by the direct search $n_M/s \lesssim
      2 \times 10^{-28}$ \cite{Ambrosio:2002qq}.  \label{fig:2PT} }
  \end{center}
\end{figure}

In Fig.~\ref{fig:2PT}, combing Eq. (\ref{eq:monopoleratio}) and
(\ref{eq:monopoannihi}), we show how magnetic monopole density to
entropy ratio depends on the critical exponent $\nu$ in 2nd order
phase transition for different critical temperatures. It is clearly
that the Parker limit \cite{Parker:1970xv} excludes the such monopoles
assuming the phase transition below the unification scale unless the
critical exponent $\nu$ is significantly above 1.  However, this is
not the case normally considered in relativistic field theories for
phase transitions.

\begin{figure}
\begin{center}
  \includegraphics[width=0.47\textwidth]{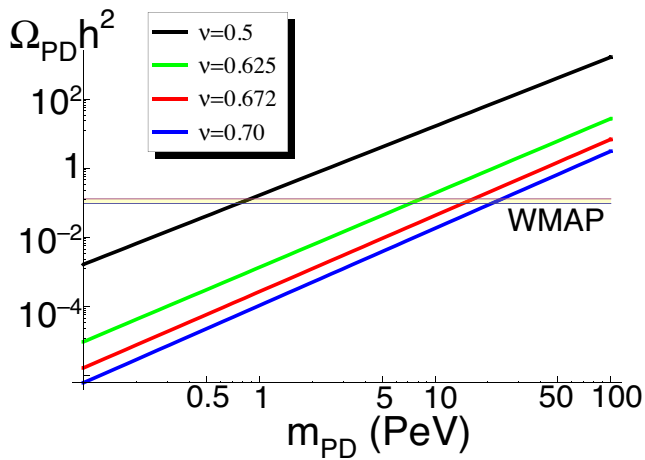}
  \caption{Relic density of topological dark matter as a function of
    its mass based on Eq. (\ref{eq:denstiy}) if it is non-thermally
    produced during a second order phase transition.  The yellow
    horizontal band denotes the relic density $0.094 < \Omega_{m}h^2 <
    0.129$ preferred by WMAP data.  We assume $x_c = m_{PD}/T_c = 50$.
\label{fig:DM} } \end{center}
\end{figure}

On the other hand, the point-like defects may be magnetic monopole
under a $U(1)$ gauge theory unrelated to electromagnetism (``hidden
$U(1)$'').  We assume that there is no corresponding hidden plasma to
dissipate the energy of $M$ and we ignore the annihilation.  We can
also ignore the annihilation for point-like defects based on global
symmetries because there is no long-range force among them.  For
non-thermal production of topological defects to dominate, they have
to be heavy enough so that they never stay in \textsl{chemical}
equilibrium once produced.  At the critical temperature, when $x_c
\equiv m_{PD}/T_c > x_f$, which ranges from 20 to 30 for different
monopole coupling strengths and phase transition temperatures, the
relics density could be derived from Eq. (\ref{eq:monopoleratio}) as
\begin{equation}
  \label{eq:denstiy}
  \Omega_{PD} h^2 \approx 1.5\times 10^{9} \left( \frac{x_c
      T_c}{1\textrm{TeV}}\right) \left( \frac{30 T_c}{M_{pl}}
  \right)^{\frac{3 \nu}{1+\nu}} \ .  
\end{equation}
If we take $x_c=50$, the relic density is a function of $T_c$, as
shown in Fig.~\ref{fig:DM}.  In order to account for the cold dark
matter abundance, we need $T_c \sim O(1)$ PeV to in the classical case
and $T_c \sim O(10)$ PeV in the typical quantum cases \footnote{For
  the charged dark matter, one has to check whether it is effectively
  collisionless \cite{Ackerman:2008gi}. Our ``hidden'' monopole is so
  heavy that its small number density makes the average time for its
  scatter greater than the age of the universe.}.

What kind of model can lead to realistic topological dark matter?  One
obvious possibility is that there is a hidden non-abelian gauge theory
whose breaking to $U(1)$ produces magnetic monopoles.  As long as the
$U(1)$ does not mix with QED, strong bounds such as Parker's limit
\cite{Parker:1970xv} do not apply. Their annihilation cross section in
the plasma is negligible.  As an example, the vector-like model of
dynamical supersymmetry breaking by Izawa--Yanagida
\cite{Izawa:1996pk} and Intriligator--Thomas
\cite{Intriligator:1996pu} has $SO(6)$ global symmetry.  Gauging
$SO(3)$ subgroup embedded diagonally into $SO(3) \times SO(3) \subset
SO(6)$, one can see that it breaks to $SO(2)$ and produces magnetic
monopoles.

It is not clear if skyrmions can be created by the same Kibble--Zurek
mechanism, as they are topologically non-trivial configurations in the
bulk rather than the boundary conditions.  However, skyrmions are
baryonic composites in the underlying gauge theory and hence may be
thermalized independent of the production mechanism.  Note that many
composite Higgs models ({\it e.g.}\/, little Higgs theories) proposed
in the literature can have skyrmions as topological solitons (see
Table~\ref{table:Models} \footnote{Although we restricted our
  consideration in $3+1$ dimensions for simplicity, skyrmion solution
  also exists in models with a compactified extra dimension. See for
  instances, Ref. \cite{Hill:2000rr}.}). Their masses are expected in
the 10~TeV region because $f_\pi \approx 1$~TeV in these theories from
the naturalness argument. Once thermally produced, the correct
abundance of topological dark matter could be obtained with a
relatively strong coupling $g_{PD} \sim 3$.  Since the global symmetry
$G$ in those models in Table~\ref{table:Models} is approximate, we may
ask whether the skyrmion is metastable.  Gauging a subgroup of $G$ may
induce the skyrmion to decay through instanton effects
\cite{D'Hoker:1983kr}. However, the enormous suppression factor
proportional to $\exp(- 8 \pi^2 /g^2)$ will make its life time much
longer than the age of our universe \cite{'tHooft:1976up} \footnote{It
  is interesting to notice that a naive estimate of the skyrmion decay
  width $\Gamma \sim f_{\pi} \exp(- 8 \pi^2 /g^2)$ will lead to the
  required life time $\tau \sim 10^{26}$s to explain the cosmic ray
  anomalies if $g \simeq 0.803$. However, details are
  beyond the scope of this Letter.}.

\begin{table}
\caption{Summary of popular composite Higgs models in $3+1$ dimensions that
  generate skyrmions. 
\label{table:Models}}
  \centering 
  \begin{tabular}{l||c|c|c}
\hline
\hline
& & & \\[-2mm]
~~~Models & $G$~  & $H$~ & $\mathcal{\pi}_3(G/H)$~~
\\[1.5mm]
\hline
\hline
& & & \\[-2.4mm]
~~~Minimal Moose \cite{Arkani-Hamed:2001nc}     & 
  $SU(3)^2$ & $SU(3)$~ & $\bf{Z}$~~    \\[1.5mm]
\hline
& & &\\[-2.4mm]
~~~Littlest Higgs \cite{Arkani-Hamed:2002qy}    & 
  $SU(5)$~ & $SO(5)$~ & $\bf{Z_2}$~~ \\[1.5mm]
\hline
& & & \\[-2.4mm]
~~~$SO(5)$ Moose \cite{Chang:2003un}        & 
  $SO(5)^2$ & $SO(5)$~ & $\bf{Z}$~~   \\[1.5mm] 
\hline
\hline
\end{tabular}
\end{table}

Let us now comment on the consequence of topological dark matter on
cosmic ray signals. For the case of skyrmion dark matter, we can
imagine that skyrmions will decay through some higher dimension
operators analogous to proton decay in Grand Unified Theory (GUT). The
most economical way is to consider GUT-suppressed dimension 6
operators, with its lifetime \cite{Arvanitaki:2008hq}
\begin{equation} 
  \tau \sim 8 \pi \frac{M_{GUT}^4}{m_{PD}^5} = 3 \times 10^{27} s \left(
    \frac{\textrm{TeV}}{m_{PD}} \right)^5 \left( \frac{M_{GUT}}{2
      \times 10^{16}\textrm{GeV}} \right)^4 \ . 
\end{equation}
The final decay products of the skyrmions would be some meson states
with extra fundamental fermions, for instance charged leptons. We can
imagine that the main branching ratio of the skyrmion decay is the one
into a light meson state below GeV. The light meson mixes with the
Higgs boson, so its coupling to the SM fermions is proportional to
their masses and will dominantly decays into $\mu$ pairs. As long as
the mass of the skyrmion is multi-TeV, the muon dominated leptonic
final state will naturally explain \cite{IMSY, Meade:2009iu} the
PAMELA excess in $e^+$ \cite{Adriani:2008zr} while no excess in
$\bar{p}$ \cite{Adriani:2008zq} as well as slow decline $E^{-3.0}$ in
the $e^-+e^+$ spectrum as reported by FERMI
\cite{Collaboration:2009zk} which steepens at about 1 TeV as measured
by H.E.S.S. \cite{Collaboration:2008aaa}.

In summary, we have considered the possibilities that point-like
topological defects, such as monopoles and skyrmions, as the viable
dark matter candidates. We apply the Kibble--Zurek mechanism to the
non-thermal production of monopoles by a second order phase
transition, and find that the abundance is much larger than the one
originally estimated by Kibble. Depending on critical exponent in the
correlation length, the hidden monopoles could account for the correct
relics density for the mass range of approximately 1--10 PeV. The
thermally produced skyrmion of mass 1--10 TeV can also provide the
correct relics density of cold dark matter, whose decay may account
for the flux and spectral shape of electrons and positrons recent
observed by PAMELA and FERMI satellites and HESS. A similar
consideration excludes any models that produce magnetic monopoles
above TeV scale but below the inflation scale, such as Pati--Salam
unification.

\acknowledgments

The authors thank Shin'ya Aoki, Koichi Hamaguchi, Tetsuo Hatsuda,
Simeon Hellerman, Joel Moore, Hai Qian, Masahito Ueda and Tsutomu
Yanagida for useful discussions.  This work was supported in part by
World Premier International Research Center Initiative, MEXT, Japan.
The work of H.M. was also supported in part by the U.S. DOE under
Contract DE-AC03-76SF00098, in part by the NSF under grant
PHY-04-57315, and in part by the Grant-in-Aid for scientific research
(C) 20540257 from Japan Society for Promotion of Science (JSPS). The
work of J.S. was also supported by the Grant-in-Aid for scientific
research (Young Scientists (B) 21740169) from JSPS.

\end{document}